\documentclass[12pt]{amsart}
\usepackage{colortbl}
\usepackage{amsmath}
\usepackage{amsxtra}
\usepackage{amscd}
\usepackage{amsthm}
\usepackage{amsfonts}
\usepackage{amssymb}
\usepackage{eucal}
\usepackage{epsfig}
\usepackage{graphics}
\def\({\left(}
\def\){\right)}

\newcommand{\nn}{\nonumber}
\newcommand{\bea}{\begin{eqnarray}}
\newcommand{\ena}{\end{eqnarray}}
\def\bel{\begin{eqnarray}}
\def\enl{\end{eqnarray}}
\newcommand{\be}{\begin{eqnarray*}}
\newcommand{\en}{\end{eqnarray*}}
\newcommand{\ba}{\begin{array}}
\newcommand{\ea}{\end{array}}

\newenvironment{tenumerate}{
  \begin{enumerate}
  
  }{\end{enumerate}}
\newcommand{\bi}{\begin{tenumerate}}
\newcommand{\ei}{\end{tenumerate}}
\newcommand{\isoto}[1][]%
{{\mathop{\buildrel{\sim}\over\longrightarrow}\limits_{#1}}}

\def\[{\left[}
\def\]{\right]}
\newcommand{\la}{\lambda}

\newcommand{\al}{\alpha}

\newcommand{\s}{\sigma}
\newcommand{\z}{\zeta}

\newcommand{\om}{{\omega}}

\numberwithin{equation}{section}

\def\half{\textstyle{\frac  1 2}}

\def\bi{\mathbf{i}}

\newcommand{\ccal}[1]{
\scalebox{1}{
$\hskip-.13cm{\operatorname{\text{\usefont{U}{BOONDOX-cal}{m}{n}#1}}}\hskip-.12cm$
}
}
\newcommand{\corr}{\scalebox{.6}{$\mathrm{corr}$}}

\begin{document}
\begin{title}[On consistency of perturbed generalised minimal models]
{On consistency of perturbed generalised minimal models}

\end{title}
\date{\today}
\author{H.~Boos and  F.~Smirnov}
\address{HB: Physics Department, University of Wuppertal, D-42097,
Wuppertal, Germany}\email{boos@physik.uni-wuppertal.de}
\address{FS\footnote{Membre du CNRS}: 
${}^{1}$ Sorbonne Universit\'e, UPMC Univ Paris 06\\ CNRS, UMR 7589, LPTHE\\F-75005, Paris, France}\email{smirnov@lpthe.jussieu.fr}

\begin{abstract}

We consider the massive {perturbation} of the Generalised Minimal Model introduced by Al. Zamolodchikov. The one-point functions in this case are supposed to be described
by certain function $\omega(\z,\xi)$. We prove that this function satisfies several
properties which are necessary for consistency of the entire procedure.

\end{abstract}

\maketitle

\section{Introduction}

Consider the Liouville model at the one hand and the minimal model (MM) of CFT on the other.
Several important for physics quantities are obtained for MM as analytical continuation
of the corresponding Liuoville ones. For the ratio of the three-point function of a descendant of the primary fields to the three-point function of the primary fields {themselves}
this is obvious since such ratio is fixed by the Ward identities. But for the three-point functions of the primary fields themselves it is strictly speaking not correct. Indeed, the
Liouville three-point function does not allow an analytical continuation with respect to
the coupling constant. What is true is that the {ratio} of two three-point functions
{ found by Dorn and Otto \cite{Dorn-Otto} and by Zamolodchikov and Zamolodchikov \cite{Zamol-Zamol}} 
shifted by a discreet set of constants (defined below) can be continued analytically
being given by products of gamma-functions. Corresponding computation was carefully done by Teschner \cite{T}. 

The latter observation led Al. Zamolodchikov \cite{AlZam} to assume that there is 
a solution of the conformal bootstrap which {differs} from the Liouville one but gives
the same ratio of the three-point functions. This solution is supposed to
be analytical in the entire complex plane of the coupling constant except 
the positive half-line which remains in Liouville disposal. Since the ratio of the
three-point functions remains the same the three point function for this Generalised
Minimal Model (GMM) differs from the Liouville one by a double-periodic function.
The latter cannot be defined when two periods are collinear which corresponds to
the Liouville case. 
In the present paper we shall talk only about the ratio of the shifted Liouville
three-point functions leaving apart Al. Zamolodchikov's double-periodic function,
their implementation and certain problems occurring in the process. 

The main subject of this paper  is the massive {deformation} of GMM  which
can be viewed as its
$\Phi_{1,3}$-perturbation of GMM (PGMM). The analogue of the three-point function for 
GMM will be the one point function on a cylinder. The asymptotical conditions 
are described by twists which can be different. This gives us three parameters:
one for the primary field (denoted $\al$ below) and two for twists in the 
asymptotics ($\kappa$ and $\kappa'$). If $\al$ and $\kappa$ are arbitrary, but
$\kappa'=\kappa+s p\al$, ($s\in \mathbb{Z}_{\ge 0}$ , $p$ parametrises the central
charge){,} we are on the save ground {provided} by the perturbation of minimal model.
It is clear that working with the PGMM
is rather bold idea, implementing it , one has at least
to verify that a number
of properties valid for the perturbation of MM allow a generalisation. 
This is the goal of the present paper. 

The final goal of our procedure is to obtain the correlation functions on 
a cylinder. This consists of two part local and global. Local serves two purposes:
with help of PCFT we construct the OPE, then we use the fermionic basis 
as target of the OPE. Precisely it means that we first write  the OPE for any
convenient set of local operators in the right hand side, which we  further rewrite
in the fermionic basis which provides a full space of local 
operators. The reason for this is that the one-point functions of the 
elements of the fermionic basis are simple, they are defined by one function
of two variables. It is important to realise that the two parts of the construction
use different sets of data: the OPE and the formulas relating the fermionic basis with
the basis of local operators are of completely local (UV) character,
oppositely, the one-point functions are constructed via global Matsubara data. In 
the present paper we shall concentrate on them.

The main part of the Matsubara data is provided by the function $\omega(\z,\xi|\al,\kappa,\kappa')$
explicitly described below. Some important properties of this function for PGMM were
conjectured in \cite{BS}. However, in this paper the conjectures were not always
clearly pointed out and missed proofs. The goal of this paper is to fill this gap.

\section{Preliminaries}

Lets us remind some general definitions. For an integrable model on a cylinder
consider the Matsubara space. 
In this paper we shall  deal with the  Bethe vectors which  depend on twist $\kappa$ and have spin $s$. 
These two usually come  together, so, for the sake of conciseness we shall often unite 
{them} into
$$\varkappa=(\kappa,s)\,.$$

{Our} final goal is the PGMM model on the cylinder. This is achieved by the scaling limit starting
from the inhomogeneous six-vertex model 
with straggling inhomogenieties $\z_0$, $\z_0^{-1}$
on the cylinder. In addition we 
insert a quasi-local operator as explained in \cite{HGSV,BS}.
So, we begin with the considering 
a finite  Matsubara spin chain of even length $\mathbf{n}$, in the scaling limit
$$\mathbf{n}\to\infty,\quad \z_0\to\infty\,,$$
in such a way that
$$2\pi MR=4\mathbf{n}\z_0^{-(p+1)}$$
is finite. 

Let us start by considering finite $\mathbf{n}$ and $\z_0$. For this case already some non-trivial properties will be elucidated. 
The $Q$-function
$$Q^-(\z|\varkappa)=\z^{-\kappa+s}\prod\limits_{\mathbf{j=1}}^{\mathbf{n}/2-s}\Bigl(1-\frac{\z^2}{\la_j^2}\Bigr)\,,$$
where $\la_j$ are Bethe roots, will be denoted by $Q(\z|\varkappa)$. 
The eigenvalue of the transfer-matrix is
$$T(\z|\varkappa)=\frac{a(\z)Q(\z q|\varkappa)+d(\z)Q(\z q^{-1}|\varkappa)}{Q(\z|\varkappa)}\,. $$

We shall consider two eigenvectors corresponding to different $\varkappa$, $\varkappa'$.
	
The ratio of the eigenvalues with different twists plays an important role
$$
\rho(\z)=\frac{T(\z|\varkappa')}{T(\z|\varkappa)}
\,.$$
Also we use the ratio of $Q$-functions
\bea
h(\z)=\frac{Q(\z|\varkappa')}{Q(\z|\varkappa)}. \label{h}
\ena
Here and later we avoid indicating the arguments $ \varkappa, \varkappa'$ if the
understanding of formulas is unambiguous.

We shall use familiar notations
\begin{align}
&\mathfrak{a}(\z)=\frac{a(\z)Q(\z q|\varkappa)}{d(\z)Q(\z q^{-1}|\varkappa)}\,,\quad
dm(\z)=\frac 1 {\rho(\z)( 1+\mathfrak{a}(\z))}\,,\nn\\&(F\star G)(\z,\xi)=\int_{\gamma}F(\z,\eta)G(\eta,\xi)dm(\eta)\,,\quad
 \delta^-_\z f(\z)=
f(\z q)-\rho(\z)f(\z)\,,\quad \nn\\&f_{\mathrm{left}}(\z,\eta)=\frac 1 {2\pi i}\delta^-_\z \psi(\z/\eta,\al)\,,\quad f_{\mathrm{right}}(\eta,\xi)=\delta^-_\xi \psi(\eta/\xi,\al)\,,\nn\\
&K_\al(\z)=\frac 1 {2\pi i} \bigl( \psi(\z q,\al)-\psi(\z q^{-1},\al)\bigr)\,.\nn
\end{align}

For the  function {$\psi(\z,\al)$} we shall take one of the following
\begin{align}
\psi_0(\z,\al)=\frac {\z^{\al}(\z^2+1)}{2(\z^2-1)},\quad \psi_+(\z,\al)=\frac {\z^{\al}}{\z^2-1}
\,.\label{psis}
\end{align}
A part of our computations does not depend on the choice, in this case we shall
use simply {$\psi(\z,\al)$}.

Our main object is the function {$\omega$} defined by
\begin{align}
\frac 1 4 \omega(\z,\xi|\al,\kappa,\kappa')&=-\bigl(f_\mathrm{left}\star f_\mathrm{right}-f_\mathrm{left}\star  
R^\mathrm{dress}_\al\star f_\mathrm{right}\bigr)(\z,\xi)\label{eqomega}\\
&+\delta^-_\z\delta^-_\xi\Delta_\z^{-1}\psi(\z/\xi,\al)\nn\,.\end{align}
 $f_\mathrm{right}$, $f_\mathrm{left}$ are constructed using $\varkappa$, $\varkappa'$ respectively, 
where Bethe roots, $\z$, $\xi$ are inside the contour which goes {counterclockwise},
and  the dressed resolvent is defined by  the equation
\begin{align}
R^\mathrm{dress}_\al+K_\al\star R^\mathrm{dress}_\al=K_\al\,,\qquad 
R^\mathrm{dress}_\al+R^\mathrm{dress}_\al\star K_\al=K_\al\,,\label{dR}
\end{align}
$\varkappa$, $\varkappa'$ enter through $\rho(\z)$.
For certain relations between the parameters the function 
{$\omega(\z,\xi|\al,\kappa,\kappa')$} happens to be independent 
{of} the choice of \eqref{psis}, we shall return to this.

The solutions to \eqref{dR} allow analytical continuation for both arguments by virtue of the equations themselves. But applying to function with singularities like $G_{\mathrm{right}}(\eta,\xi)$,
for example, we have to be careful about the position of the singularity
($\xi$ in this case). Historically, we put it inside the contour. In the present context it
is convenient to use more natural from general point of view choice putting the 
singularities outside of the contour. Then the formula for $\omega(\z,\xi)$ 
\footnote{

In some cases when this will not cause any confusion,
we will omit the dependence of $\om$ on $\alpha,\kappa,\kappa' $} 
undergoes
some modification.

The main part is 
\begin{align}
\bigl(f_\mathrm{left}\star G_{\mathrm{right}}\bigr)(\z,\xi)
\end{align}
where we introduced
\begin{align}\label{eqG}
G_{\mathrm{right}}(\eta,\xi)=
f_\mathrm{right}(\eta,\xi)-\bigl(K_\al\star G_{\mathrm{right}}\bigr)(\eta,\xi)\,.
\end{align}
The first goal is to switch to the contour $\gamma_B$ which contains the Bethe roots only:
\begin{align}
&G_{\mathrm{right}}(\eta,\xi)+\int\limits_{\gamma_B}K_\al(\eta/\mu)G_{\mathrm{right}}(\mu,\xi)dm(\mu)\nn\\&=f_\mathrm{right}(\eta,\xi)+2\pi i K_\al(\eta/\xi)\frac 1 {1+\mathfrak{a}(\xi)}=V_{\mathrm{right}}(\eta,\xi)\nn\,
\end{align}
where we defined
\begin{align}&V_{\mathrm{right}}(\eta,\xi)=\frac 1 {1+\mathfrak{a}(\xi)}{\psi(q\eta/\xi,\al)}
+\frac 1 {1+\bar{\mathfrak{a}}(\xi)}{\psi(q^{-1}\eta/\xi ,\al)}-\rho(\xi)\psi(\eta/\xi ,\al)\nn\\
&=\frac 1 {Q(\xi|\varkappa)Q(\xi|\varkappa')T(\xi|\varkappa)}\bigl(d(\xi)U_{\mathrm{right}}(\eta,\xi)-a(\xi)U_{\mathrm{right}}(\eta ,\xi q)\bigr)\,,\nn\end{align}
with
\begin{align}
&U_{\mathrm{right}}(\eta,\xi)=Q(\xi q^{-1}|\varkappa)Q(\xi|\varkappa')\psi(q\eta/\xi)-Q(\xi |\varkappa)Q(\xi q^{-1}|\varkappa')\psi(\eta/\xi)\,.\nn
\end{align}

Consider
\begin{align}
&\int\limits _{\gamma}f_\mathrm{left}(\z,\eta)G_{\mathrm{right}}(\eta,\xi)dm(\eta)=\int\limits _{\gamma _B}f_\mathrm{left}(\z,\eta)G_{\mathrm{right}}(\eta,\xi)dm(\eta)\nn\\
&
+G_{\mathrm{right}}(\z,\xi)\frac 1{1+\mathfrak{a}(\z)}-
{2\pi i}f_{\mathrm{left}}(\z,\xi)
\frac 1{1+\mathfrak{a}(\xi)}\,.\nn
\end{align}
Let
\begin{align}
&f_\mathrm{left}(\z,\eta)=V_\mathrm{left}(\z,\eta)+\frac 1{1+\mathfrak{a}(\z)}K_\al(\z/\eta)\,,\nn
\end{align}
where
\begin{align}
&V_\mathrm{left}(\z,\eta)={\frac{1}{2\pi i}}
{\biggl(}\frac 1{1+\bar{\mathfrak{a}}(\z)}\psi(q\z/\eta,\al)
+\frac 1{1+\mathfrak{a}(\z)}\psi(q^{-1}\z/\eta,\al)
-\rho(\z)
\psi(\z/\eta,\al)
{\biggr)}
\nn\\&=\frac 1 {Q(\z|\varkappa)Q(\z|\varkappa')T(\z|\varkappa)}\bigl(d(\z)U_{\mathrm{left}}(\z,\eta)-a(\z)U_{\mathrm{left}}(q\z, \eta )\bigr)\,,\nn\end{align}
with
\begin{align}
	&U_{\mathrm{left}}(\z,\eta)=
	\frac{1}{2\pi i}\bigl(Q(\z q^{-1}|\varkappa)Q(\z|\varkappa')\psi( q^{-1}\z/\eta)-Q(\z |\varkappa)Q(\z q^{-1}|\varkappa')
	\psi(\z/\eta)\bigr)\,.\nn
	\end{align}

Then
\begin{align}
&\int\limits _{\gamma}f_\mathrm{left}(\z,\eta)G_{\mathrm{right}}(\eta,\xi)dm(\eta)=\int\limits _{\gamma_ B}V_\mathrm{left}(\z,\eta)G_{\mathrm{right}}(\eta,\xi)dm(\eta)\nn\\
&+\frac 1{1+{\mathfrak{a}}(\z)}V_{\mathrm{right}}(\z,\xi)-
{2\pi i}f_{\mathrm{left}}(\z,\xi)
\frac 1{1+\mathfrak{a}(\xi)}\nn\\&=
\int\limits _{\gamma_ B}V_\mathrm{left}(\z,\eta)G_{\mathrm{right}}(\eta,\xi)dm(\eta)
+W(\z,\xi)\,,\nn
\end{align}
where
\begin{align}
&W(\z,\xi)=
\frac 1 {(1+{\mathfrak{a}}(\z))(1+\bar{\mathfrak{a}}(\xi))}\psi(q^{-1}\z/\xi,\al)
-\frac 1 {(1+\bar{\mathfrak{a}}(\z))(1+{\mathfrak{a}}(\xi))}\psi(q\z/\xi,\al)\nn\\
&+\Bigl(\frac {\rho(\z)} {1+{\mathfrak{a}}(\xi)}-\frac {\rho(\xi)} {1+{\mathfrak{a}}(\z)}\Bigr)\psi(\z/\xi,\al)\,.\nn
\end{align}

Finally
\begin{align}
\frac 1 4 \omega(\z,\xi|\al)&=-\bigl(V_\mathrm{left}\star V_\mathrm{right}-V_\mathrm{left}\star  
R^\mathrm{dress}_\al\star V_\mathrm{right}\bigr)(\z,\xi)\label{eqomega}\\
&+\delta^-_\z\delta^-_\xi\Delta_\z^{-1}\psi(\z/\xi,\al)-W(\z,\xi)\nn\,,\end{align}
where {the convolution $\star$ is now defined via the integration contour 
$\gamma_B$ that goes only around the Bethe-roots $\la_k$ with the points
$\z$, $\xi$  taken {\it outside} the contour $\gamma_B$.}

\section{$\psi_0$ and $\psi_+$}

Let us begin with the papers \cite{HGS}.
There the fermionic basis was {used} to
obtain the expectation values of 
spinless quasi-local operators (local operator $\mathcal{O}$
on the lattice with the left ``tale" of $q^{\al\sigma_j^3}$). 
In this case or the function
$\omega(\z,\xi|\al,\kappa,\kappa')$  the parameters are restricted:
$$\kappa'=\kappa+\al\,.$$
In \cite{HGSIII} $\omega(\z,\xi|\al,\kappa,\kappa')$ was defined as
deformed canonical second kind Abelian differential.

Later in \cite{HGSIV,HGSV} the restriction was eased. In these papers we allowed
local operators to have positive spin $\s_\mathcal{O}$ and, similarly to well-known CFT construction,
compensated it by introducing ``screening
operators". Then the expectation values {were} expressed
via 
$\omega(\z,\xi|\al,\kappa,\kappa')$ with
\begin{align}\kappa'=\kappa+\al+s_\mathcal{O}\,,\label{eman}\end{align}
$$ s_\mathcal{O}=s'-s\,.$$
So, we have certain freedom in the choice of $\kappa'$, we called this ``emancipation" of
$\kappa'$. The ``emancipation" was not complete, the complete one is achieved in the GMM.

The definition in terms of the deformed Abelian integrals is beautiful, but it is difficult
to use it for the scaling limit. Hence we were forced to use another, equivalent,
definition which used the integral equations described above. The equivalence is
proved in \cite{BoGo}. The paper \cite{BoGo} uses the function $\psi_0$, but it proves that
for $\kappa'=\kappa+\al\,.$ this function can {be} changed by 
$\psi_+$. This looks as a technical detail, but in fact it has important consequences for
the scaling limit improving the convergence of integrals for $R^{\mathrm{dress}}$,
$G_{\mathrm{right}}$ and $\omega$ in the range $0\le \al \le 2$ which we
consider as the fundamental domain. {We shall return to this point later.}

Our first goal is to generalise results of \cite{BoGo} proving that
$\psi_0$ and $\psi_+$ provide the same final result, i.e. the same $\omega$
for the ``emancipated" case \eqref{eman}.  { Below we will write 
the subscript `+' for all functions calculated at $\psi=\psi_+$ as, for instance, 
we denote the function $\omega$ taken with $\psi=\psi_+$ by $\omega_+$ .
}

Recall  the relation \cite{BoGo}
\begin{align}
\bigl(h(\eta q^{-1})-h(\eta q)  \bigr)dm(\eta)=\Bigl( {h(\eta)}-\frac 1 {\rho(\eta)}h(\eta q)
\Bigr)\frac {d\eta^2}{\eta^2}\,,\label{etaeta}
\end{align}
Suppose that 
$$s'\ge s\,.$$

{Let us first take  $\psi$ equal to $\psi_0$.}
Introduce the function $G_{\mathrm{left}}$ by the equation
$$ G_{\mathrm{left}}(\z,\xi)+\bigl(G_{\mathrm{left}}\star K_\al\bigr)(\z,\xi)=f_{\mathrm{left}}
(\z,\xi)\,.$$

We are interested in the behaviour of this function for $\z\to 0$.
Define 
{
$$
g_{\mathrm{left}}(\xi)=\mathop{\lim}_{\z\to 0}\z^{-\al}G_{\mathrm{left}}(\z,\xi)\,.
$$
}
It is easy to see that $g_{\mathrm{left}}$ satisfies the equation
\begin{align}
g_{\mathrm{left}}(\xi)+\int\limits_{\gamma_B}g_{\mathrm{left}}(\eta)K_{\al}(\eta/\xi)
{dm(\eta)}
=
-{{\xi}^{-\al}}
\frac{q^{-\al}(q^{\al}-q^{\kappa+\kappa'-s-s'})(q^{\al}-q^{\kappa-\kappa'-s+s'})}{{4\pi i}(1+q^{2(\kappa-s)})}\label{eqg}
\end{align}
Slightly {generalising}  \cite{BoGo},  we claim that for  the function
$ g_{\mathrm{left}}(\xi)$ can be found ``explicitly" for
\begin{align}
\al=\kappa'-\kappa-s'+s\,,\quad s'\ge s\,,
\label{domain}
\end{align}
The solution in this domain is
\begin{align}
 g_{\mathrm{left}}(\xi)=\frac 1 C
\frac{q^{-\al}(q^{\al}-q^{\kappa+\kappa'-s-s'})(q^{\al}-q^{\kappa-\kappa'-s+s'})}{{4\pi i}(1+q^{2(\kappa-s)})}
\bigl(h(\xi q^{-1})-h(\xi q)
\bigr)\,,\label{solg}\end{align}
where $h$ has been defined in \eqref{h}, and
$$
C=\frac 1 2(q^{\al}-q^{-\al})\Bigl({-1}+\delta_{s,s'}\Bigl(\prod _{j=1}^{{\mathbf{n}/2}-s}\frac {\la_j}{\la'_j}\Bigr)^2\Bigr)\,.
$$
Let us comment briefly on the derivation. Using \eqref{etaeta} under the assumption
\begin{align}
\al-\kappa'+\kappa\in\mathbb{Z}\,,\label{assump}
\end{align}
we {compute}
\begin{align}&\frac 1 {2\pi i  }
\int\limits_{\gamma_B}\bigl(h(\eta q^{-1})-h(\eta q)\bigr)\psi(\eta/\sigma,\al)dm(\eta)=
{\frac 1 {2\pi i  }}\int\limits_{\gamma_B}h(\eta)\psi(\eta/\sigma,\al)\frac{d\eta^2}{\eta^2}\nn\\
&=-h(\sigma)-\bigl({\mathop{\lim}_{\eta^2\to 0}+\mathop{\lim}_{\eta^2\to\infty}}\bigr)
h(\eta)\psi(\eta/\sigma,\al)\,.\nn
\end{align}
Now it is clear that for \eqref{domain} the function $ g_{\mathrm{left}}(\xi)$ \eqref{solg} solves the 
equation \eqref{eqg}. 

Using the symmetry
$G_{\mathrm{left}}\star f_{\mathrm{right}}
=f_{\mathrm{left}}\star G_{\mathrm{right}}\,,$
one computes
\begin{align}
{\frac 1 {2\pi i  }}\int\limits_{\gamma}
\eta^{-\al}G_{\mathrm{right}}(\eta,\xi)dm(\eta)=
\frac 1 {q^{\al}-q^{-\al}}\xi^{-\al}(q^{-\al}-\rho(\xi))\,.
\label{eqGright}
\end{align}
{ 
As byproduct we also derive that
$$
\lim_{\z\to 0} \z^{-\al}G_{\mathrm{right}}(\z,\xi)=0.
$$

Now we are ready to calculate the difference between $\om$ taken at $\psi=\psi_0$ and 
$\om$ taken at $\psi=\psi_+$ that we denoted $\om_+$. 
First, we observe that because of (\ref{eqGright}), the difference $G_{\mathrm{right}}-
G_{+,\mathrm{right}}$ 
satisfies the integral equation  
$$
G_{\mathrm{right}}-
G_{+,\mathrm{right}}=-K_{+,\al}\star(G_{\mathrm{right}}-
G_{+,\mathrm{right}})
$$
that gives us
$$
G_{\mathrm{right}}=G_{+,\mathrm{right}}.
$$
Then 
\bea
&&
\frac14\bigl(\om(\z,\xi)-\om_+(\z,\xi)\bigr)\nn\\
&&
=-\Bigl((f_{\mathrm{left}}-f_{+,\mathrm{left}})\star G_{\mathrm{right}}\Bigr)(\z,\xi)+
\delta^-_\z\delta^-_\xi\Delta_\z^{-1}(\psi_0(\z/\xi,\al)-\psi_+(\z/\xi,\al))\nn\\
&&
=-\frac{\z^{\al}}{4\pi i}\bigl(q^{\al}-\rho(\z)\bigr)
\int\limits_{\gamma}\eta^{-\al} G_{\mathrm{right}}(\eta,\xi)dm(\eta)+
\frac{(q^{\al}-\rho(\z))(q^{-\al}-\rho(\xi))}{2(q^{\al}-q^{-\al})}\bigl(\z/\xi\bigr)^{\al}
=0\nn
\ena
where again we used (\ref{eqGright}). So, we obtain the equivalence between $\omega$'s computed by \eqref{eqomega} with $\psi=\psi_0$ and 
$\psi=\psi_+$.
}

\section{Symmetry $\varkappa\leftrightarrow \varkappa'$}

The $P$-symmetry 
requires the symmetry under the switch of $\varkappa$ and $\varkappa'$. Let us prove that this takes {place} even for finite $\mathbf{n}$. 

I this section we shall show that for $\psi=\psi_+$ the symmetry holds
\begin{align}
\omega(\z,\xi|\varkappa,\varkappa',\al)=\rho(\z)\rho(\xi)\omega(\z,\xi|\varkappa',\varkappa,\al)
\end{align}

Our goal is to evaluate the difference
\begin{align}
\mathrm{diff}(\z,\xi)=\frac 1 4 \bigl(\omega(\z,\xi|\varkappa,\varkappa',\al)-\rho(\z)\rho(\xi)\omega(\z,\xi|\varkappa',\varkappa,\al)\bigr)\,.\nn
\end{align}

Define
$$\omega_0(\z,\xi|\varkappa,\varkappa')=\delta^-_\z\delta^-_\xi
\Delta_\z^{-1}\psi(\z/\xi,\al)\,,$$
where the dependece on $\varkappa,\varkappa'$, enters  implicitly through $\rho$. 
It is easy to see that
\begin{align}
\omega_0(\z,\xi|\varkappa,\varkappa')-\omega_0(\z,\xi|\varkappa',\varkappa)=(\rho(\z)-\rho(\xi))\psi(\z/\xi,\al)\,.\nn
\end{align}
Further
	\begin{align}
	&\omega_0(\z,\xi|\varkappa,\varkappa')-W(\z,\xi|\varkappa,\varkappa')-\rho(\z)\rho(\xi)\bigl( \omega_0(\z,\xi|\varkappa',\varkappa)-W(\z,\xi|\varkappa',\varkappa)\bigr)\nn\\&=\frac {d(\z)d(\xi)Z(\z,\xi)-d(\z)a(\xi)Z(\z,\xi q)-a(\z)d(\xi)Z(\z q,\xi)+
		a(\z)a(\xi)Z(\z q,\xi q)} {T(\z|\varkappa)T(\xi|\varkappa)Q(\z|\varkappa)Q(\xi|\varkappa)Q(\z|\varkappa')Q(\xi|\varkappa')}\,,\nn\end{align}
where
\begin{align}
	Z(\z,\xi)&=\bigl(Q(\xi|\varkappa')Q(q^{-1}\xi|\varkappa)Q(\z|\varkappa)Q(q^{-1}\z|\varkappa')\nn\\&-
	Q(\xi|\varkappa)Q(q^{-1}\xi|\varkappa')Q(\z|\varkappa')Q(q^{-1}\z|\varkappa)
	\bigr)\psi(\z/\xi,\al)\,.\nn
	\end{align}
This allows to compute the difference which we 
 write down explicitly
  \begin{align}
&\mathrm{diff}(\z,\xi)\nn\\&=\frac {d(\z)d(\xi)X(\z,\xi)-d(\z)a(\xi)X(\z,\xi q)-a(\z)d(\xi)X(\z q,\xi)+a(\z)a(\xi)X(\z q,\xi q)} {T(\z|\varkappa)T(\xi|\varkappa)Q(\z|\varkappa)Q(\xi|\varkappa)Q(\z|\varkappa')Q(\xi|\varkappa')}\,,\nn
\end{align}
where
\begin{align}
&X(\z,\xi)=I(\z,\xi|\varkappa,\varkappa')-I(\z,\xi|\varkappa',\varkappa)+Z(\z,\xi)\,,\nn\\
&I(\z,\xi|\varkappa,\varkappa')=-\int\limits_{\gamma_B}U_{\mathrm{left}}(\z,\eta|\varkappa,\varkappa')U_{\mathrm{right}}(\eta,\xi|\varkappa,\varkappa')dm(\eta,\varkappa)\nn\\
&+\int\limits_{\gamma_B}\int\limits_{\gamma_B}U_{\mathrm{left}}(\z,\eta|\varkappa,\varkappa')R_{\mathrm{dress}}(\eta,\mu|\varkappa,\varkappa')
U_{\mathrm{right}}(\mu,\xi|\varkappa,\varkappa')dm(\eta,\varkappa)dm(\mu,\varkappa)\,.\nn
\end{align}
It will be convenient to use the less symmetric form
\begin{align}
&I(\z,\xi|\varkappa,\varkappa')=-\int\limits_{\gamma_B}U_{\mathrm{left}}(\z,\eta|\varkappa,\varkappa')
H_{\mathrm{right}}(\eta,\xi|\varkappa,\varkappa')dm(\eta,\varkappa)\nn\\
&H_{\mathrm{right}}(\eta,\xi|\varkappa,\varkappa')\nn\\&\qquad =U_{\mathrm{right}}(\eta,\xi|\varkappa,\varkappa')-\int\limits_{\gamma_B}R_{\mathrm{dress}}(
\eta,\mu|\varkappa,\varkappa')U_{\mathrm{right}}(\mu,\xi|\varkappa,\varkappa')dm(\mu,\varkappa)\,.\nn
\end{align}

We claim that the function $X(\z,\xi)$ is of the form
\begin{align}
X(\z,\xi)=(\z/\xi)^\al\mathrm{Pol}(\z^2,\xi^2)\,,
\end{align}
with $\mathrm{Pol}(\z^2,\xi^2)$ being a polynomial .
For $Z(\z,\xi)$ this is obvious. Consider $X(\z,\xi|\varkappa,\varkappa')$ as function of $\z$. Clearly, it 
is $(\z/\xi)^\al$ multiplied by a rational function of $\z^2$. The latter might have singularity at
the pinch of the contour 
by either $\eta^2-\z^2$  or $\eta^2q^2-\z^2$
with { the poles coming from} $1/Q(\eta|\varkappa)$. 
However, in these cases the integrand contains the multipliers
$Q(\zeta|\varkappa)$ or  $Q(\zeta q^{-1}|\varkappa)$
which cancel the singularities.

The degrees of the polynomial depend on the choice of the function $\psi$ above. Namely.
\begin{align}
&\psi=\psi_0: \quad 
\,{\mathrm{deg}}_{\z^2}\mathrm{Pol}(\z^2,\xi^2)=\mathbf{n}-s-s'\,,\ \ 
\ \ \ \ \ \,{\mathrm{deg}}_{\xi^2}\mathrm{Pol}(\z^2,\xi^2)=\mathbf{n}-s-s'\,,\label{degs}\\
&\psi=\psi_+: \quad 
{\mathrm{deg}}_{\z^2}\mathrm{Pol}(\z^2,\xi^2)=\mathbf{n}-s-s'-1\,,\ \ 
{\mathrm{deg}}_{\xi^2}\mathrm{Pol}(\z^2,\xi^2)=\mathbf{n}-s-s'\,.\nn
\end{align}

So, our goal is to show that the polynomial $\mathrm{Pol}_{\mathbf{n}}(\z^2,\xi^2)$ 
vanishes. 
Consider the combination
\begin{align}
&Y(\z,\xi)=Q(q\z|\varkappa)X(\z,\xi)+Q(q^{-1}\z|\varkappa)X(\z q,\xi)\,,\nn
\end{align}
It is convenient to split $X(\z,\xi)$, and correspondingly,  $Y(\z,\xi)$ into two parts
\begin{align}
&X_1(\z,\xi)=I(\z,\xi|\varkappa,\varkappa')+Z(\z,\xi)\nn\\
&X_2(\z,\xi)=-I(\z,\xi|\varkappa',\varkappa)\nn\,,\\
&Y_i(\z,\xi)=Q(q\z|\varkappa)X_i(\z,\xi)+Q(q^{-1}\z|\varkappa)X_i(\z q,\xi)\,,\quad i=1,2\,.\nn
\end{align}

Consider $Y_1(\z,\xi)$.
We have
		\begin{align}
	&Q(\z q|\varkappa)U_{\mathrm{left}}(\z,\eta|\varkappa,\varkappa')+Q(q^{-1}\z|\varkappa)U_{\mathrm{left}}(\eta,q\z|\varkappa,\varkappa')\label{X1}\\&
	=-Q(\z, \varkappa') Q(q^{-1}\z, \varkappa) Q(q\z, \varkappa)\Bigl(K_\al(\z,\eta)
	+\frac{1}{2\pi i}\psi(\z/\eta,\al) \frac 1 {h(\z)} \bigl( h(q^{-1}\z)-h(q\z)\bigr)\Bigr)\,,\nn
	\end{align}
 where we used notation (\ref{h}).

Recall  the relation {(\ref{etaeta})}.
In  the convolution
\begin{align}
\int\limits_{\gamma_B}K_\al(\z,\eta)H_{\mathrm{right}}(\eta,\xi)dm(\eta)=-H_{\mathrm{right}}(\z,\xi)+U_{\mathrm{right}}(\z,\xi)\,.\nn
\end{align}
set $\z=\la_\mathbf{m}$.
It is easy to see form \eqref{etaeta} that
\begin{align}
\mathop{\lim}_{\z\to\la_{\mathbf{m}}}
\frac 1 {h(\z)} \bigl( h(q^{-1}\z)-h(q\z)\bigr)\int\limits_{\gamma_B}\psi(\z/\eta,\al)H_{\mathrm{right}}(\eta,\xi)dm(\eta)=H_{\mathrm{right}}(\la_\mathbf{m},\xi)\,.\nn
\end{align}
Indeed, the integrand is pinched by two poles producing a pole at $\z=\la_{\mathbf{m}}$
which simplifies with the zero of $1/{h(\z)} $. 
Now by simple algebra we compute
\begin{align}
Y_1(\la_\mathbf{m},\xi)&=Q(\la_\mathbf{m}, \varkappa') Q(q^{-1}\la_\mathbf{m}, \varkappa) Q(q\la_\mathbf{m}, \varkappa)U_{\mathrm{right}}(\la_{\mathbf{m}},\xi)\nn\\&
+Q(\la_\mathbf{m} q|\varkappa)Z(\la_\mathbf{m},\xi)+Q(\la_\mathbf{m} q^{-1}|\varkappa)Z(\la_\mathbf{m} q,\xi)=0\,.\nn
\end{align}
Another  combination to consider is  $Y_2(\z,\xi)$.
Here the  computation is simple. We have
\begin{align}
&Q(\z q|\varkappa)U_{\mathrm{left}}(\z,\eta|\varkappa',\varkappa)+Q(q^{-1}\z|\varkappa)U_{\mathrm{left}}(\eta,q\z|\varkappa',\varkappa)\label{X2}\\&
=Q(\z, \varkappa) \bigl(\psi(q ^{-1}\z/\eta) Q(q^{-1}\z, \varkappa') Q(q \z, \varkappa) -
  \psi(q\z/\eta)Q (q^{-1}\z, \varkappa)Q(q \z, \varkappa')\bigr)\,,\nn
\end{align}
which implies that
$$Y_2(\la_\mathbf{m},\xi)=0\,,$$
for trivial reasonsince $Q(\la_\mathbf{m}, \varkappa)=0$.

So, we come to the conclusion that 
\begin{align}
X(\la_{\mathbf{m}},\xi)=0,\quad \mathbf{m}=\mathbf{1},\cdots ,\mathbf{n}/2-s\,.\nn
\end{align}
Similarly one proves that
\begin{align}
X(\la'_{\mathbf{m}},\xi)=0,\quad \mathbf{m}=\mathbf{1},\cdots ,\mathbf{n}/2-s'\,.\nn
\end{align}
Due to \eqref{degs} we see that  sufficient to assert the statement for $\psi=\psi_+$. 
In the next section we explain why this will be our preferred case for the scaling limit.

\section{Shift of $\al$}

Here we consider the scaling limit which describes the massive model.
This model in general corresponds to the perturbation of the GMM, for a discreet set of parameter $\al$ it  is equivalent to
the sG model. 

It will be convenient to switch to the  variables
\begin{align} 
Z=\z^{\frac 1 {p+1}}\,,\quad X=\xi^{\frac 1 {p+1}}\,.\label{changevars}
\end{align}
Then the DDV equations are turn into
\begin{align}
&\log\mathfrak{A}(Z,\kappa)=\pi i  MR(Z-Z^{-1})-\frac{4\pi i }  {p }\kappa \\&
-\int_{0}^{\infty}
G(Z/Z')\Bigl(\log(1+\mathfrak{A}(Z'e^{i0},\kappa))-
\log(1+\overline{\mathfrak{A}(Z' e^{i0},\kappa)}\Bigr)\frac{dZ'}{Z'}\,,\nn
\end{align}
where $M$ is the soliton mass, and  the kernel is 
\begin{align}
G(Z)=\int_{-\infty}^{\infty}\frac{\sinh(\frac {\pi k} 2(1-p))}
{4\pi \sinh(\frac {\pi k} 2p)\cosh\(\frac {\pi k} 2\)}Z^{i k} dk
\,.
\end{align}

First, we would like to introduce convenient notations. 

Allover this section we are in the  region $0<p<1/2$ and  $\al$ is small:
\begin{align}
\al<1\,,\label{rangeal}
\end{align}
in order that $\al+2p<2$.
Recall
\begin{align}
&
t_a(\al)=
\half \cot{\textstyle \frac \pi {2}}(\al+a (p+1))\,,\nn
\end{align}
We define further
$$R(Z)=\rho(\z)\,,$$
which leads to  the asymptorics
\begin{align}
\log R(Z)\mathop{\simeq}_{Z\to\infty} \  \sum_{j=1}^\infty
Z^{-(2j-1)}\varDelta I_{2j-1}\,,\quad
\log R(Z)\mathop{\simeq}_{Z\to 0} \  \sum_{j=1}^\infty
Z^{(2j-1)}\varDelta \bar I_{2j-1}\,.\nn
\end{align}
In the CFT limit
$$\varDelta I_{2j-1}=
M^{-(2j-1)}C_{2j-1}(p)(I_{2j-1}(\kappa')-I_{2j-1}(\kappa))\,,$$
$I_{2j-1}(\kappa)$ being the eigenvalue of the  local integral of motion, to compare  them with the  BLZ one uses
$$p=\frac{\beta^2}{1-\beta^2}\,,\quad \kappa=-\frac P {2(1-\beta^2)}\,.$$

The kernel $K_\al$ turns into
$$\mathcal{K}_\al(Z)=\frac 2 {p+1}K_\al(\z)\,,$$
the multiplier is introduced in order to take care of change of integration variables under
the integral. $K_\al$ is defined via  $\psi=\psi_+$.
We  have chosen  $\psi_+(\z,\al)$ here because with this choice the
integral converges well for functions which do not grow at $\z=0,\infty$. This
is important got the scaling limit because the Bethe roots fill the entire half-axis
$(0,\infty)$ of  $\z$.

We have the integral operator 
on the half-line with the kernel
$$(\mathcal{K}_\al f)(Z)=\int\limits_0^{\infty}\mathcal{K}_\al (Z/S)f(S)\frac{dS}S\,.$$
  Notice that 
 \begin{align}\mathcal{K}_\al(1/Z)= \mathcal{K}_{2-\al}(Z)\,, \label{transK}\end{align}
 so, $\mathcal{K}_{2-\al}$ gives the kernel of the transposed operator. 
 
 Now we perform the usual modification of the integral equation deriving the
 DDV equations
 
 \noindent 1.
First convolution
\begin{align}
&(F\circ G)(Z,X)=\int_0^\infty F(Z,U)G(U,X)d\mu(U)\,,\quad d\mu(U)=\frac {dU}{UR(U)}
\,.\nn
\end{align}

\noindent
The `bare' resolvent is defined via
\begin{align}{\mathcal{R}}_{\al}-{\mathcal{K}}_{\al}\circ {\mathcal{R}}_{\al}={\mathcal{K}}_{\al}\,.\label{eq}\end{align}

 \noindent 2.
Second convolution
\begin{align}
&
(F * G)(Z,X)=\int_0^\infty F(Z,U)G(U,X)dm(U)\,,\nn \\
&
\quad dm(U)=\frac {dU}{UR(U)}\Biggl(\frac  1 {1+\overline{\mathfrak{A}(Ze^{i0})}} +
\frac  1 {1+{\mathfrak{A}}(Ze^{-i0})} 
\Biggr)\,,\nn
\end{align}
the interval {$(0,\infty)$} contains the locus of Bethe roots. 

\noindent
Dressed resolvent
\bea
\mathcal{R}^{\mathrm{dress}}_{\al}+\mathcal{R}^{\mathrm{dress}}_{\al} * \mathcal{R}_{\al}=\mathcal{R}_{\al}\,.\label{Rdress}
\ena

\noindent
3. Bra and ket
\begin{align}
F_\mathrm{left}=f_\mathrm{left}+f_\mathrm{left}\circ \mathcal{R}_{\al}\,,
\quad F_\mathrm{right}=f_\mathrm{right}+\mathcal{R}_{\al}\circ f_\mathrm{right}\,.
\label{defFlr}
\end{align}

\noindent 4.
Function $\Omega$ is defined by the change of variables
$$\Omega(Z,X)=\frac 1{2(p+1)}{\omega(\z,\xi)}\,.$$
It has a splitted form
\begin{align}
&\Omega(Z,X|\al)=\Omega ^{(1)}(Z,X|\al)+\Omega ^{(2)}(Z,X|\al)\,,\label{omega12}\\
&\Omega ^{(1)}(Z,X|\al)=\bigl(-F_\mathrm{left}*F_\mathrm{right}+F_\mathrm{left}*\mathcal{R}^{\mathrm{dress}}_{\al}*F_\mathrm{right}
\bigr)(Z,X)\,,\nn\\
&\Omega ^{(2)}(Z,X|\al)=\bigl(f_\mathrm{left}\circ F_\mathrm{right}\bigr)(Z,X)+\Omega _0(Z,X|\al)\,.\nn
\end{align}
where
$\Omega _0$ is obtained from $\frac 1{2(p+1)}\delta^-_\z\delta^-_\xi\Delta_\z^{-1}\psi(\z/\xi,\al)$ by change of variables.  Notice that the shift $\z\to \z q$ becomes $Z\to Z e^{\pi i}$.
  
The function $S^k$ is a formal eigenfunction of $I-\mathcal{K}_\al$ 
(when the integral converges) with the eigenvalue 
\begin{align}
1-\int\limits_{0}^{\infty}\mathcal{K}_\al(S)S^{k-1}{dS}= \frac{2 \sin\frac{\pi}{2} (pk-\al ) \cos(\frac{\pi} 2k)}
    { \sin\frac{\pi}{2}(k(p+1)- \al)}\,.\label{eigen}
\end{align}

For imaginary $k$ it is a true eigenfunction of continuous spectrum. 
As usual one can 
regularise the integral out of the convergency region taking \eqref{eigen} for definition.

We continue with the following simple observation
 \begin{align}
 \mathcal{K}_{\al+2p}(Z)={Z^2\bigl( \mathcal{K}_\al(Z)+a\cdot
 Z^{-\frac{(2-\al)}{p+1}}\bigr)}
 \,,\label{shiftK}
\end{align}
where
$$a={\textstyle \frac{2}{\pi(p+1)}}\sin{\textstyle \frac\pi {p+1}}(2-\al)\,.$$
Rewrite  \eqref{shiftK} as
\begin{align}
& \mathcal{K}_{\al+2p}=U^2 \widetilde{ \mathcal{K}}_{\al+2p}U^{-2}\,,
\nn\\
&\widetilde{ \mathcal{K}}_{\al}= \mathcal{K}_{\al}+a\cdot g_{\al}^-\otimes  g_\al^+{}^t\,,\nn
\end{align}
where $U$ is the operator of multiplication by $Z$, and
$$g_\al^\pm(Z)={Z^{\pm \frac{(2-\al)}{p+1}}}\,.$$

We shall need the resolvent
\begin{align}\widetilde{\mathcal{R}}_{\al}-\widetilde{\mathcal{K}}_{\al}\circ \widetilde{\mathcal{R}}_{\al}=\widetilde{\mathcal{K}}_{\al}\,.\label{eqtilde}\end{align}

The operator 
$\widetilde{\mathcal{K}}_\al $ looks as a perturbation of $\mathcal{K}(\al )$ by one-dimensional projector.
This is not quite the case because the functions $g^\pm_\al(Z)$ do not
belong to $L_2$. Still it is true that the resolvent  $\widetilde {\mathcal{R}}_\al $ differs from the unperturbed one satisfying (\ref{eq})
via
$$
\widetilde{\mathcal{R}}_\al =\mathcal{R}_\al +b\cdot G^-_\al\otimes  G^+_\al\,,
$$
with $ G^\mp_\al$ being the solutions to the equations
$$G^-_\al-\mathcal{K}_\al\circ G^-_\al=c^-\cdot
g_\al^-\,,\quad G^+_\al- G^+_\al\circ \mathcal{K}(\al)=c^+\cdot
g^+_\al\,,$$
$b,c^\pm$ are some constants depending on $\al$.
It will be often  convenient to rewrite using \eqref{transK} the latter equation as
$$ G^+_\al-\mathcal{K}_{2-\al} \circ G^+_\al=c^+\cdot
g^+_\al\,.$$

Let us consider first the simplest case $R(Z)=1$. In that case 
the resolvent is found by means of Fourier transform:
\begin{align}
&\mathcal{R}_\al (Z,Z')=V_\al(Z/Z')\,,\nn\\
&V_\al(Z)=\int_{-\infty}^{\infty}Z^{i x}\frac{\sinh\frac\pi 2\left((1-p)x-i\al\right)}
{4\pi\sinh\frac\pi 2\left(px+i\al\right)\cosh\left(\frac{\pi x} 2\right)}dx\,.\nn
\end{align}
One easily finds
$$G^\pm_\al(Z)=Z^{\pm 1}\,,$$
and 
\begin{align}
b=\frac 1{2 \pi  t_1(\al)}\,.\label{b}
\end{align}
The function $V_\al(Z)$  will be actively used in what follows.

We claim that generally $G^\pm_\al(Z)$ grow for $\log Z\to\pm\infty$ as $O(Z^{\pm})$
rapidly decreasing for $\log Z\to\mp\infty$.
This is
consistent with convergence of the integrals above because $\al<1$ in our domain \eqref{rangeal}. 
We shall return to the constant $b$ soon.
Let us fix the normalisation requiring that
\begin{align}G^\pm_\al(Z)=Z^{\pm 1}\bigl(1+o(1)\bigr)\,,\quad
\quad\log Z\to \pm\infty\,.\label{normG}\end{align}
We  have
\begin{align}R(Z)=1+O(Z^{\mp 1}) \,,\quad\log Z \to \pm \infty\,.\label{asR}\end{align}
Write \eqref{eq}
explicitly
\begin{align}\mathcal{R}_\al(Z,X)-\int\limits_{0}^{\infty}\mathcal{K}_\al(Z/S)\mathcal{R}_\al(S,X)
\frac{dS}{SR(S)}=\mathcal{K}_\al(Z/X)\,.
\end{align}
Shift the arguments
$$S=S'e^{-a},\ \ Z=Z'e^{-a}\,,\ \ X=X'e^{-a}\,,$$
and send $a$ to $\infty$. Clearly the asymptotics of $\mathcal{R}_\al(Z/X)$ in this limit
coincides with   the resolvent for $R(Z)=1$. 
Same is true for $\widetilde{\mathcal{R}}_\al(Z/X)$. 
Comparing with the case  $R(Z)=1$ one easily proves that the constant $b$ does not depend on $R(Z)$
as far as \eqref{asR} takes place.

Let us emphasise an  unusual feature of the perturbation containing the growing functions $g^\pm_\al(Z)$.

Introduce 
\begin{align}
&F_\al(S,X)-
\int_{0}^{\infty}\mathcal{K}_\al(S/S')F_\al(S',X)\frac{dS'}{S'R(S')}\label{eqFright}
=\ccal{f}_{\mathrm{right}}(S,X)\,.
\end{align}
For the variables $X,\ Z$ we have the convention:
$$-\pi<\arg{X}<0\,,\quad -\pi<\arg{Z}<0\,.$$

In order to solve the equation \eqref{eqFright} efficiently we divide  $F_\al(S,X)$ into two pieces
\begin{align}
F_\al(S,X)=F^{(0)}(S,X)+F_\al^{\corr}(S,X)\,,\quad
F^{(0)}(S,X)=-\frac {2SX(R(S)R(X))^{1/2}}{(S^2-X^2)}\,,\label{defFcorr}
\end{align}
and invert an operator with difference kernel  arriving at the equation with compact kernel for $F_\al^{\corr}(S,X)$
\begin{align}
&F^{\corr}_\al(S,X)-\int_{0}^{\infty}V_\al(S/S')\Bigl(\frac 1{R(S')}-1\Bigr)F^{\corr}_\al(S',X)
\frac{dS'}{S'}\label{eqFcorr}\\&
=\Psi^-_\al(S/X)
-\Psi^+_\al(S/X)R(X)
\nn\\&-F^{(0)}(S,X)+\int_{0}^{\infty}V_\al(S/S')\Bigl(\frac 1{R(S')}-1\Bigr)F^{(0)}(S',X)
\frac{dS'}{S'}\,,\nn
\end{align}
where
\begin{align}
&\Psi^{+}_\al(S)=\int_{0}^{\infty}S^{i x}\frac{\exp\left(\frac\pi 2\left((p+1)x+i\al\right)\right)}
{4i  \sinh\frac\pi 2\left(px+i\al\right)\cosh\left(\frac{\pi x} 2\right)}dx\,,\nn\\
&\Psi^{-}_\al(S)=\Psi^{+}_\al(S)-\frac {2S}{S^2-1}
\,.\nn
\end{align}

Another advantage of \eqref{defFcorr} is that it clearly implies the  standard asymptotics in $Z$:\footnote{Following \cite{BS} we call
the asymptotical expantion for $\log Z\to\pm\infty$ of the form $\sum_{j=1}^{\infty}f_{2j-1}Z^{\mp(2j-1)}$ the ``standard asymptotics"}
\begin{align}
F_\al(S,X)\mathop{\simeq} _{\log X\to\pm \infty} R(X)^{\frac 1 2}\Bigl(F_{\al,\pm 1}(\theta)X^{\mp 1}+F_{\al,\pm 3}X^{\mp 3}+\cdots\Bigr)\,.\label{as}\end{align}
This follows from the observation that the right hand side of \eqref{eqFcorr}, multiplied by $R(X)^{-1/2}$ , has standard asymptotics in 
$X$ and the coefficients
of the asymptotical series are rapidly decaying for at $\pm\infty$ functions of $\log S$.

The following limits exist
\begin{align}&H(X|\al)=2 t_1(\al)\mathop{\lim}_{\log S\to\infty}SF_\al(S,X)\,.\nn\\
&H^\dag(Z|\al)=2t_1(\al)\mathop{\lim}_{\log S\to-\infty}S^{-1}F_{2-\al}(S,Z)\,,\nn\end{align}
where the coefficient is introduced for further convenience. 
After some consideration we conclude that
$$G^-_\al(S)=-\frac 1 2F_{\al,-1}(S)\,,\qquad G^+_{\al}(S)=-\frac {2t_1(\al)}{H(\al)} F_{2-\al,1}(S)\,,$$
where
\begin{align}
H(\al)=\mathop{\lim}_{X\to 0}X^{-1}H(X|\al)
\,.\nn
\end{align}

The formulas above lead to the first important result
\begin{align}
&U^{-2}\mathcal{R}_{\al+2p}U^{2} =\mathcal{R}_\al+
\frac {F_{\al,-1}\otimes F_{2-\al,1}}{2\pi H(\al)}\,.\label{mainR}
\end{align}
From this equation we derive in particular
\begin{align}
&\left(\frac X S\right)^2 F_{\al+2p}(S,X)
 -F_\al(S,X)+\frac {H(X|\al)}
{H(\al)}F_{\al,-1}(S)=0\,.\label{FFF}
\end{align}
This identity has the corollary
\begin{align}
\lim_{S\to 0}\frac{X^2} SF_{\al+2p}(S,X|\al)=2\frac {H(X|\al)}
{H(\al)}
\label{corr}
\end{align}

Now we use the `dressed' resolvent $\mathcal{R}_\al^{\scalebox{.6}{$\mathrm{dress}$}}$ 
defined in (\ref{Rdress})
and introduce
\begin{align}
F_{\al}^{\scalebox{.6}{$\mathrm{dress}$}}=F_{\al}-\mathcal{R}_\al^{\scalebox{.6}{$\mathrm{dress}$}}*
F_{\al}\,,\label{Rdr}
\end{align}
which solves the equation 
\begin{align}
F_{\al}^{\scalebox{.6}{$\mathrm{dress}$}}+\mathcal{R}_\al* F_{\al}^{\scalebox{.6}{$\mathrm{dress}$}} =F_{\al}\,.\nn
\end{align}
Summing up a geometrical progression one derives
\begin{align}
 &U^{-2}\mathcal{R}^{\scalebox{.6}{$\mathrm{dress}$}}_{\al+2p}U^2
=\mathcal{R}^{\scalebox{.6}{$\mathrm{dress}$}}_\al-\frac{1} {2\pi i\bigl({\Omega^{(1)}_{1,-1}(\al)+iH(\al)}\bigr)}
 F^{\scalebox{.6}{$\mathrm{dress}$}}_{\al,-1}
\otimes
F^{\scalebox{.6}{$\mathrm{dress}$}}_{2-\al,1}\label{main}
\,,
\end{align}
where
$$
\Omega^{(1)}_{1,-1}(\al)=-\frac 1 {2\pi i}\Bigl(
F_{2-\al,1}* F_{\al,-1}-F_{2-\al,1}*\mathcal{R}^{\scalebox{.6}{$\mathrm{dress}$}}_\al*F_{\al,-1}
\Bigr)\,,
$$
is a particular asymptotical coefficient of the function $\Omega^{(1)}(Z,X|\al)$ which is one of the main
goals of our consideration.

According to our previous results the function $\Omega(Z,X|\al)$ consists of two parts:
\begin{align}
\Omega(Z,X|\al)=\Omega^{(1)}(Z,X|\al)+\Omega^{(2)}(Z,X|\al)\,.\label{Om=Om1+Om2}
\end{align}
The first term is
\begin{align}
\Omega^{(1)}(Z,X|\al)=-\frac 1 {2\pi i}\Bigl(
F_{2-\al}(Z)* F_{\al}(X)-F_{2-\al}(X)*\mathcal{R}^{\scalebox{.6}{$\mathrm{dress}$}}_\al*F_{\al}(X)
\Bigr)\,,\label{defom1}
\end{align}
where we omit the first arguments of $F_{2-\al}$, $F_\al$ which is the integration variable. 
Notice that
$$\Omega^{(1)}(Z,X|\al)=-\frac 1 {2\pi i} F_{2-\al}(Z)*F^{\scalebox{.6}{$\mathrm{dress}$}}_{\al}(X)=-\frac 1 {2\pi i} F^{\scalebox{.6}{$\mathrm{dress}$}}_{2-\al}(X)*F_{\al}(X)\,.$$
It is immediate to derive for $\Omega^{(1)}(Z,X|\al)$
\begin{align}
\left(\frac{X}{Z}\right)^2\Omega^{(1)}&(Z,X|\al+2p)-\Omega^{(1)}(Z,X|\al)+i\frac{H^\dag(Z|\al)H(X|\al)}{H(\al)}
\label{final1}\\&+
\frac{\left(\Omega^{(1)}_{-1}(Z|2-\al)-iH^\dag(Z|\al)\right)
\left(\Omega^{(1)}_1(X|\al)+iH(X|\al)\right)}
{\Omega^{(1)}_{1,-1}(\al)+iH(\al)}=0\,.\nn
\end{align}

The study  of the function $\Omega^{(2)}(Z,X|\al)$ is more complicated. We have
\begin{align}
&\Omega^{(2)}(Z,X|\al)=U(Z,X) -{\frac 1 {\pi i}}
\int\limits_{0}^\infty U(Z,S)
F_\al
(S,X)\frac{dS}{SR(S)}\,,\label{om2}
\end{align}
with
\begin{align}
U(Z,X)&=\frac 1 4\Bigl((1-R(Z))(1+R(X))\frac{Z^2+X^2}{Z^2-X^2}\nn\\&-
(1+R(Z))(1-R(X))\frac{2ZX}{Z^2-X^2}\Bigr)\,.\nn
\end{align}

We have the asymptotics
\begin{align}
\Omega(Z,X|\al)\mathop {\simeq}_
{ \log Z\to \pm\infty}R(Z)^{1/2}\sum\limits_{j=1}^\infty
Z^{\mp(2j-1)}\Omega_{\pm(2j-1)}(X|\al)\,,\nn
\end{align}
and
\begin{align}
\Omega_{2j-1}(X|\al)\mathop{\simeq}_{\log X \to \pm\infty}R(X)^{1/2}\sum\limits_{j=1}^\infty
X^{\mp(2j-1)}\Omega_{2j-1,\pm(2k-1)}(\al)\,,\nn
\end{align}
and similarly for $\Omega^{(1)}$, $\Omega^{(2)}$. This asymptotical behaviour is obvious for 
 $\Omega^{(1)}$ while for $\Omega^{(2)}$ it is a result of more subtle analysis.

 Let us compute $\Omega^{(2)}_1(X|\al)$. By definition we have
\begin{align}
\Omega^{(2)}_1(X|\al)&=-\frac 1 {\pi i}\int\limits_{-0}^{\infty}S\left(1-\frac 1 {R(S)} \right)F_\al(S,X)\frac{d S}{S}\nn\\
&+\frac 1 {4\pi i}\varDelta I_{1}\int\limits_{0}^{\infty}\left(1+\frac 1 {R(S)} \right)F_\al(S,X)\frac{d S}{S}\nn\\
&-\frac 1 4\Bigl(4X(1-R(X))+\varDelta I_{1}(1+R(X))
\Bigr)\,,\nn
\end{align}
The second integral can be computed
\begin{align}
\frac 1 {4\pi i}\int\limits_{0}^{\infty}\left(1+\frac 1 {R(S)} \right)F_\al(S,X)\frac{d S}{S}=\frac 1 4
\Bigl((1+R(X))+2it_0(\al)(1-R(X))\label{2int}
\Bigr)\,,
\end{align}
which allows to simplify
\begin{align}
\Omega^{(2)}_1(X|\al)&=-\frac 1 {\pi i}\int\limits_{0}^{\infty}S \left(1-\frac 1 {R(S)} \right)F_\al(S,X)\frac{d S}{S}\nn\\
&-\frac 1 2\Bigl(2X-it_0(\al)\varDelta I_{1}\Bigr)(1-R(X))\,.\nn
\end{align}
On the other hand, rewrite the equation \eqref{eqFcorr} as
\begin{align}
&F_\al(S,X)=\int_{0}^{\infty}V_\al(S/S')\Bigl(\frac 1{R(S')}-1\Bigr)F_\al(S',X)
\frac{d S'}{S'}\nn\\&
+\Psi^-_\al(S/X)
-\Psi^+_\al(S/X)R(X)\,,
\nn
\end{align}
and consider the asymptotics for $S\to\infty$. The leading  term is computed replacing $V_\al$ and
$\Psi^-_\al$ by the first terms of their asymptotics. This simple computation leads to
\begin{align}
\Omega^{(2)}_1(X|\al)+\Xi(X|\al)=iH(X|\al)\,,\label{tau}
\end{align}
where
\begin{align}
\Xi(X|\al)=\frac 1 {2i}\Bigl(4t_1(\al)X(1+R(X))+t_0(\al)\varDelta I_{1}(1-R(X))
\Bigr)\nn\,.
\end{align}
Similarly
\begin{align}
\Omega^{(2)}_{-1}(Z|2-\al)+\Xi^{\,\dag}(Z|\al)=-iH^\dag(Z|\al)\,,\label{sigma}
\end{align}
where
\begin{align}
\Xi^\dag(Z|\al)=\frac 1 {2i}\Bigl(4t_1(\al)Z^{-1}(1+R(Z))+t_0(\al)\varDelta \bar{I}_{1}(1-R(Z))
\Bigr)\nn\,.
\end{align}
Introduce the function of two variables which will be important in what follows
 \begin{align}
\Xi(Z,X|\al) &= \frac 1 {2i} \Bigl\{2 t_1(\al)
    \,\frac X Z(1 + R(Z))(1 + R(X)) \nn\\&- \Bigl( \Bigl(\frac X Z \Bigr)^2 t_2(\al) + t_0(\al)\Bigr)(1 -R(Z))(1 - R(X))\Bigr\}\,.\nn
     \end{align}
The functions $\Xi(X|\al)$, $\Xi^\dag(Z|\al)$ are obtained as leading asymptotics os $\Xi(Z,X|\al)$  \begin{align}
&\Xi(Z,X|\al) \mathop{\simeq}_{Z\to \infty} Z^{-1}X\Xi(X|\al)+\cdots\,,\nn\\
&\Xi(Z,X|\al)\ \mathop{\simeq}_{ X\to 0} \, X \Xi^\dag(Z|\al)+\cdots\,,\nn
\end{align}
We shall also need the constant
\begin{align}
\Xi(\al)=\frac 1 {2i}\Bigl(8t_1(\al)-\varDelta \bar{I}_{1}\varDelta{I}_{1}t_0(\al)\Bigr)\,,\nn
\end{align}
obtained as the double asymptotics
$$\Xi(Z,X|\al) \mathop{\simeq}_{{ \log Z\to \infty}\atop{ \log X\to -\infty}} XZ^{-1}\  \Xi(\al)+\cdots\,.$$

Let us multiply \eqref{FFF} by $$\frac 1 {2\pi }\frac{U(Z,S)}{R(S)}\,,$$
and integrate over $S$. Here one has to be careful: integrating term by term we would
find divergencies at $S\to 0$ 
in the first and the last terms which cancel each other. So, let us proceed carefully. Begin from  the following identity:
$$
U(Z,S)=\Big(\frac{Z}S \Big)^2U(Z,S)-\frac14(1-R(Z))(1+R(S))\Big(1+\Big(\frac{Z}S \Big)^2\Big)+\frac12(1+R(Z))(1-R(S))\frac{Z}S
$$
Then using (5.17) after some algebra we get
\bea
&&
U(Z,S)F_{\al+2p}(S,X)=\Big(\frac{Z}X \Big)^2 U(Z,S)F_{\al}(S,X)-\Big(\frac{Z}X \Big)^2\frac14(1-R(Z))(1+R(S))F_{\al}(S,X)
\nn\\
&&-
\frac14(1-R(Z))(1+R(S))F_{\al+2p}(S,X)+\frac{Z}{X^2}\frac12(1+R(Z))(1-R(S))F_{\al}(S,X)\nn\\
&&
-\frac{H(X|\al)}{H(\al)}\Big(\Big(\frac{Z}X \Big)^2 U'(Z,S)F_{\al,-1}(S)+\frac{Z}{X^2}\frac12(1+R(Z))(1-R(S))F_{\al,-1}(S)\Big)
\nn
\ena
where
$$
U'(Z,S)=\frac12(1-R(Z))(1+R(S))\frac{S^2}{S^2-Z^2}-\frac12(1+R(Z))(1-R(S))\frac{Z S}{S^2-Z^2}\,.
$$
Hence the integral 
$$
\int_{0}^{\infty}\frac{dS}{SR(S)}U(Z,S)F_{\al+2p}(S,X)
$$
can be calculated using (5.25) and 
$$
\frac1{\pi}\int_{0}^{\infty}\frac{dS}{SR(S)}U'(Z,S)F_{\al,-1}(S)=-H^{\dag}(Z|\al)+2Z^{-1}t_1(\al)(1+R(Z))
$$


Combine all pieces and recall that $\Omega^{(2)}(Z,X|\al)$ contains additional $\frac {1} {4}U(Z,X)$, then after some calculation one arrives at
\begin{align}
& \Bigl(\frac X Z\Bigr)^2\Omega^{(2)}(Z,X|\al+2p) -\Omega^{(2)}(Z,X|\al)
-i
  \frac{H^\dag(Z|\al)H(X|\al)}{H(\al)
  }=\Xi(Z,X|\al)\label{zeroT}\,,
  \end{align}
  
It remains to add \eqref{final1} with \eqref{zeroT} and to use \eqref{tau} and \eqref{sigma}. Thus we arrive at our final result:
\begin{align}
& \Bigl(\frac X Z\Bigr)^2\Omega(Z,X|\al+2p)-\Omega(Z,X|\al)
\nn\\&+
\frac{\bigl(\Omega_{-1}(Z|2-\al)+\Xi^{\dag}(Z|\al)\bigr)
\left(\Omega_1(X|\al)+\Xi(X|\al)\right)}
{\Omega_{1,-1}(\al)+\Xi(\al)}=\Xi(Z,X|\al)\,.\label{alphashift}
\end{align}
An important feature of this identity is that apart of the explicit functions $\Xi(X|\al)$, $\Xi^{\dag}(Z|\al)$,
$\Xi(Z,X|\al)$, $\Xi(\al)$ it contains the full function $\Omega$ only. 

\section{Conclusion}

In this paper we have tried to fill several gaps in our understanding of the function 
$\omega$ which is important for the scaling limits of the six-vertex model that can give both
CFT and the sine-Gordon model. In a similar way as Al. Zamolodchikov ``emancipated''
the minimal CFT models to the Generalized Minimal Models, we tried to ``emancipate''
the definition of the function $\omega$ to the general case of three parameters 
$\al, \kappa, \kappa'$. It made sence to unite the two twist parameters with the
spins $s, s'$ of corresponding eigenvectors of the transfer matrices:
$\varkappa=(\kappa, s),  \varkappa'=(\kappa', s')$.
We defined the domain (\ref{domain}) where the functions $\omega$ enjoys 
two reflection symmetries: $\al\to 2-\al$ and $\al\to -\al$. This fact was effectively
used in paper \cite{Sm-Neg} to identify the fermionic basis with the Virasoro module
and also in our paper \cite{BS} where we incorporated the action of the integrals 
of motion into the fermionic basis.

The other important property of the function $\omega$ that we called `$P$-symmetry' 
is the symmetry with respect to the exchange $\varkappa\leftrightarrow\varkappa'$. 
This property was observed in  \cite{BS} by means of numerical 
computations. In this paper we discussed its proof that was possible to find even in case of the finite number $\mathbf{n}$ of lattice sites in Matsubara direction. 

The last outcome of this paper is the formula  (\ref{alphashift}) that relates the 
function $\omega$ at $\alpha$ to $\omega$ with the shifted $\alpha$. In fact,
this result is a generalisation of the  identity (9.11) of the paper \cite{HGSV} 
derived for $\rho=1$ to the case $\rho\ne 1$.

\section{Acknowledgements}
The authors would like to thank F. G{\"o}hmann  for stimulating discussions. 
HB acknowledges financial support by the DFG in the framework of the
research unit FOR 2316. FS is grateful to Galileo Galilei Institute where the main part of this work was done for warm hospitality.

\end{document}